\documentclass[twocolumn,preprintnumbers,superscriptaddress,nofootinbib,aps,prd,floatfix]{revtex4}
\usepackage{graphicx}
\usepackage{amssymb}
\usepackage{amsmath}
\usepackage{amsfonts}
\usepackage{xcolor}
\usepackage{url}
\usepackage{float}
\usepackage[normalem]{ulem}
\usepackage{hyperref}
\hypersetup{
  colorlinks   = true,
  urlcolor     = blue,
  linkcolor    = black,
  citecolor   = black
}
\usepackage[caption=false]{subfig}
\usepackage{verbatim}
\usepackage{tabularx,booktabs}
\newcolumntype{C}{>{$}c<{$}}
\AtBeginDocument{%
  \heavyrulewidth=.08em
  \lightrulewidth=.05em
  \cmidrulewidth=.03em
  \belowrulesep=.65ex
  \belowbottomsep=0pt
  \aboverulesep=.4ex
  \abovetopsep=0pt
  \cmidrulesep=\doublerulesep
  \cmidrulekern=.5em
  \defaultaddspace=.5em
}

\newcommand\refeq[1]{Eq.~(\ref{#1})}
\newcommand\refeqs[1]{Eqs.~(\ref{#1})}

\newcommand\citere[1]{Ref.~\cite{#1}}
\newcommand\citeres[1]{Refs.~\cite{#1}}
\def\reffi#1{\mbox{Fig.~\ref{#1}}}

\newcommand{\BR}{\text{BR}}
\newcommand{\SM}{\text{SM}}

\newcommand{\kZon}{\kappa_{Z\text{,on}}}
\newcommand{\kZoff}{\kappa_{Z\text{,off}}}

\newcommand{\kgon}{\kappa_{g\text{,on}}}
\newcommand{\kgoff}{\kappa_{g\text{,off}}}
\newcommand{\kion}{\kappa_{i\text{,on}}}
\newcommand{\kioff}{\kappa_{i\text{,off}}}

\newcommand{\muOff}{\mu_\text{off}}
\newcommand{\GamRat}{\Gamma^H / \Gamma^H_\text{SM}}

\begin{document}
\providecommand{\abs}[1]{\lvert#1\rvert}
\preprint{DESY-26-008}

\title{On the robustness of the indirect determination of the 
\\[.3em]width of the detected Higgs boson
}
\begin{abstract}
The indirect determination of the total width of the detected Higgs boson that is carried out by the experimental collaborations at the LHC relies on the assumption that the coupling modifiers for the on-shell and off-shell couplings are the same. However, physics beyond the Standard Model affecting the on-shell and off-shell regions differently could invalidate this assumption, 
so that the actual width of the detected Higgs boson could be larger than the bounds obtained under this assumption. 
Relaxing the assumption and investigating different types of extensions of the Standard Model, we analyse under which conditions a larger total width of the detected Higgs boson is compatible with all experimental and theoretical constraints. For the considered scenarios of
scalar extensions with an additional state contributing as a resonance or at 
the loop
level,
we find that the indirect bounds obtained by ATLAS and CMS remain valid over large parts of the parameter space, with the exception of parameter regions where the additional 
particles
have
relatively small masses. 
We discuss the potential of experimental searches for new 
particles
to further constrain such scenarios. Based on the existing experimental and theoretical constraints we conclude that relaxing the assumption of equal on-shell and off-shell coupling modifiers that is used in the experimental analyses at the LHC yields an upper bound on the total width of the detected Higgs boson in realistic extensions of the Standard Model that is only weakened by up to a factor of about two compared to the case where this assumption is valid.
\end{abstract}

\author{Panagiotis Stylianou}\email{stylianou.panagiotis.1@ucy.ac.cy} 
\affiliation{
Department of Physics, University of Cyprus, P.O.\ Box 20537, 1678 Nicosia, Cyprus\\[0.1cm]}
\author{Georg Weiglein} \email{georg.weiglein@desy.de}
\affiliation{Deutsches Elektronen-Synchrotron DESY, Notkestr.~85,  22607  Hamburg,  Germany\\[0.1cm]}
\affiliation{II.\  Institut f\"ur  Theoretische  Physik, Universit\"at  Hamburg, Luruper Chaussee 149, 22761 Hamburg, Germany\\[0.1cm]}

\maketitle

\section{Introduction}
\label{sec:int}

The discovery of a 
Higgs boson with a mass of about $125$~GeV playing a central role for electroweak symmetry breaking has initiated continued efforts to precisely measure its properties, in particular the couplings to Standard Model (SM) particles~\cite{Duhrssen:2004cv} as well as its mass and width. The possibility of additional couplings to exotic states that are potentially undetectable is only weakly constrained by the upper limit of the total width obtained by direct measurements by ATLAS and CMS~\cite{ATLAS:2014euz,CMS:2024eka} which 
is affected by the limited
mass resolution. The CMS direct limit of $\GamRat< 330$~MeV ($95\%$ CL)~\cite{CMS-PAS-HIG-21-019,CMS:2024eka} 
is orders of magnitude larger than the SM prediction of $4.1$~MeV~\cite{LHCHiggsCrossSectionWorkingGroup:2016ypw} and leaves room for exotic decays.
A high-precision direct measurement of the total Higgs-boson width would only be possible with future collider concepts \cite{FCC:2018byv,Han:2012rb,Forslund:2023reu,LinearColliderVision:2025hlt,deBlas:2025gyz}.

Several ways to indirectly determine the total Higgs-boson width have been proposed,
which can lead to much stronger bounds 
than the direct limit.
The sizeable interference between Higgs production and the continuum in the diphoton final state~\cite{Dixon:2003yb} leads to different effects which can be exploited for measurements. On the one hand a mass shift is induced~\cite{Martin:2012xc,Dixon:2013haa} that depends on the Higgs width, which has been addressed in an  
ATLAS study~\cite{ATLAS:2016kvj}. 
On the other hand the different 
dependence of the resonant and interference parts of the $\gamma \gamma$ channel with respect to the 
width~\cite{Campbell:2017rke} can be exploited, 
as it was recently done by CMS~\cite{CMS:2025zue}.
The sensitivity of such approaches is however 
limited by
the resolution of the experiments in 
comparison to the small ratio of width over mass of the Higgs boson.

By far the most sensitive indirect determination of the total width is obtained from the $g g \rightarrow H \rightarrow 4 \ell$ process for on-shell and off-shell production of the Higgs boson, making use of the different dependence of the two types of processes on the total width.
As pointed out by \citere{Caola:2013yja} (see also \citeres{Campbell:2013una,Campbell:2013wga}), while the measurement of the on-shell signal strength depends on the width,
\begin{align}
	\label{eq:muon}
        \mu_\text{on}(g g \to H \to Z Z) &= \frac{\sigma(g g \to H) \BR (H \to Z Z)}{\sigma_\SM(g g \to H) \BR_\SM (H \to Z Z)} \nonumber \\ 
        &\approx \frac{\kgon^2 \kZon^2}{\GamRat}\;,
\end{align}
this is, in good approximation, not the case for off-shell measurements,
\begin{equation}
	\begin{split}
	\label{eq:muoff}
		 \mu_\text{off}(g g \to H \to Z Z) &= \frac{\sigma(g g \to H \to Z Z)}{\sigma_\SM(g g \to H \to Z Z)} \\ 
         &\approx\, {\kgoff^2 \kZoff^2}\;.
	\end{split}
\end{equation}

\begin{figure}[t!]
	\centering 
	\includegraphics[width=7.4cm]{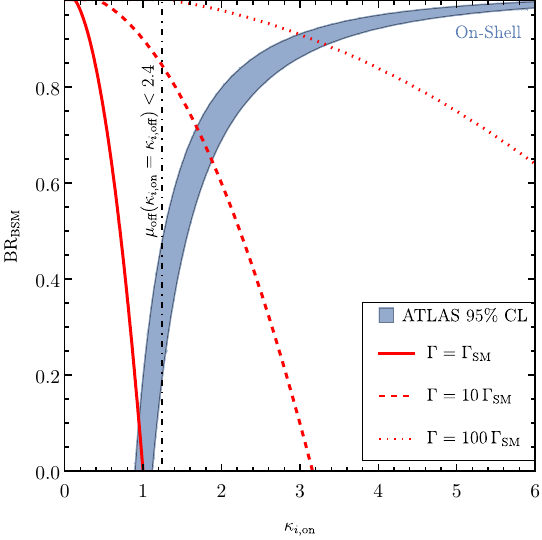}
	\caption{The blue region indicates the flat direction of the on-shell signal strength measurements for a universal Higgs coupling modifier, $\kappa \equiv \kappa_{i,\text{on}}$, using the on-shell $95\%$~CL ATLAS results on the inclusive signal strength, $\mu_ {\text{on}} = 1.01^{+0.23} _ {-0.20} $, from the four-lepton channel~\cite{ATLAS:2020rej} 
    and a varying branching ratio to (potentially undetectable) new-physics final states, $\text{BR}_{\text{BSM}}$. 
    The red solid, dashed and dotted lines indicate different values for the total Higgs width, and the black dot-dashed line is the off-shell limit $\mu_\text{off} \simeq \kappa^4 < 2.4$ under the assumption $\kappa_{i,\text{on}} = \kappa_{i,\text{off}} = \kappa$.
    \label{fig:flatdir}}
\end{figure}

In fact, a determination of the Higgs-boson width that is only based on on-shell results as the one in \refeq{eq:muon} suffers from the ``flat
direction'' where the on-shell signal strength $\mu_\text{on}$ remains unaffected by an appropriate simultaneous increase of both $\GamRat$ and $\kion = \kappa_\text{univ}$, where 
$\kappa_\text{univ} = \kappa$ is a universal coupling modifier. This is
demonstrated in Fig.~\ref{fig:flatdir}~(see also \citeres{Bechtle:2014ewa,Azatov:2022kbs}).

On the other hand, the total width can be determined from the measurements of the on-shell and off-shell signal strengths 
given in \refeq{eq:muon} and \refeq{eq:muoff} without encountering a flat direction under 
the assumption that the on-shell coupling modifiers are equal to their off-shell counterparts, $\kion = \kioff$.
In this way ATLAS and CMS 
have obtained results for the total Higgs-boson width 
having smaller uncertainties
than the results based on the methods mentioned 
above~\cite{CMS:2022ley,ATLAS:2023dnm,CMS:2024eka,ATLAS:2024jry}. The strictest limits from ATLAS and CMS on the total Higgs-boson width obtained indirectly are
\begin{align}
    \Gamma_H &= {4.5}^{+3.3} _ {-2.5} \quad (\text{ATLAS}\;\cite{ATLAS:2023dnm}), \nonumber\\
    \Gamma_H &= {3.2}^{+2.4}_{-1.7} \quad\;\; (\text{CMS}\;\cite{CMS:2022ley}).
    \label{eq:expres}
\end{align}

However, throughout this work, we will instead use the conservative upper limits on the 
Higgs-boson width and the off-shell signal strength
at the 95\% CL
from ATLAS~\cite{ATLAS:2023dnm,ATLAS:2024jry}%
\footnote{We use the ATLAS limit from the histogram approach instead of the neural simulation-based inference as this enables a clear definition of the off-shell signal region in terms of the invariant mass of the reconstructed $Z$-pair. We additionally note that the observed limit from CMS is 
stronger
than the expected limit.}
\begin{equation}
	\label{eq:atlaslim}
    \GamRat < 2.5 \quad \text{and} \quad
	\muOff < 2.4 
    \;.
\end{equation}

Apart from the $ZZ$ channel, the off-shell production of $W^+W^- \rightarrow \ell^+\ell^- \nu \bar{\nu}$ has also been considered by ATLAS and is consistent with the SM with reduced sensitivity~\cite{ATLAS:2025okx}. Furthermore, the possibility of an off-shell measurement in the $t\bar{t}t\bar{t}$ final state provides the unique opportunity to obtain a measurement of the Higgs-boson width (using similar assumptions as the $VV$ channels) without the dependence on (effective) gauge--Higgs couplings~\cite{Cao:2016wib,Cao:2019ygh}. This has also been explored experimentally by ATLAS~\cite{ATLAS:2024mhs},
but the current sensitivity is significantly lower than for the $ZZ$ channel.

In general, new-physics effects could impact the off-shell region differently than the on-shell region, 
requiring care when interpreting the results~\cite{Logan:2014ppa,Englert:2014aca,Englert:2014ffa,Cacciapaglia:2014rla,Gainer:2014hha,Azatov:2014jga,Buschmann:2014sia,Anisha:2025nhr}. 
Given the consistency of the current measurements in the off-shell region with the SM expectations, 
instead of relying on assumptions on the Higgs couplings one can directly interpret the off-shell measurements as a bound on new-physics effects arising from concrete UV-models,
see e.g.\ \citere{Englert:2014ffa}, or on
higher-dimensional 
operators within Effective Field Theory frameworks~\cite{Gainer:2014hha,Azatov:2014jga,Cacciapaglia:2014rla,Corbett:2015ksa,Englert:2017aqb,Goncalves:2020vyn,Anisha:2024xxc}.
While those bounds are of interest in their own right, they do not lead to direct conclusions regarding the reliability of the indirect determination of the total Higgs-boson width based on the assumption of equal on-shell and off-shell coupling modifiers.

In the present paper we address the question of how robust  
the indirect determination of the total Higgs-boson width using the measurements of the on-shell signal rates together with the off-shell results and in particular from the $ZZ$ channel carried out by ATLAS and CMS is 
in view of the fact that it is based on the theoretical assumption that the on-shell coupling modifiers are equal to their off-shell counterparts. In fact, as indicated by \refeqs{eq:expres} and (\ref{eq:atlaslim}), this indirect determination has meanwhile reached impressively small uncertainties, which emphasizes the need to assess its robustness. 
Relaxing the assumption
about the equality of the on-shell and off-shell coupling modifiers, we 
investigate how much the bounds on the total Higgs-boson width could degrade in realistic scenarios of physics beyond the SM (BSM) in view of the existing theoretical and experimental constraints, where the latter arise both from search limits and from precision measurements.

As discussed above the on-shell measurements alone are insensitive to a ``flat direction'' in which a significant enhancement of the total width is possible in connection with a significant increase of a universal on-shell coupling modifier $\kappa_\text{on}$. Since the experimental results for the off-shell signal strength are 
restricted to be relatively close to the SM prediction, 
values of the total width significantly exceeding the bound obtained under the assumption of the equality of the on-shell and off-shell coupling modifiers 
as given in \refeq{eq:atlaslim} 
can only be accommodated in view of the existing experimental results if the off-shell contributions are reduced in comparison to the on-shell coupling modifiers $\kappa_\text{on}$.
This constraint imposes important restrictions on the type of BSM physics contributions that 
could give rise to a significant enhancement of the total Higgs-boson width
while being compatible with the experimental data.

Furthermore, unitarity constraints restrict the possibilities for $\kappa^{\text{on}} > 1$. Thus, BSM physics giving rise to a significant enhancement of the total width of the detected Higgs boson beyond the bound of \refeq{eq:atlaslim} can only be compatible with the existing experimental and theoretical constraints if it 
enhances $\kappa_\text{on}$ while respecting the unitarity constraints and yielding
a reduction of the off-shell contributions.
An increased value of the Higgs coupling to vector bosons, $\kappa_{V, \text{on}}$, needs to be compensated in the longitudinal vector boson scattering process $VV \rightarrow VV$ by effects of additional BSM states. 
This limits the possible options from a model-building perspective~\cite{Azatov:2022kbs,Forslund:2023reu} to theories such as the generalised Georgi-Machacek models%
\footnote{While we refer here specifically to models that 
can be extended to high scales, we note that sigma models with a non-compact group symmetry, 
interpreted as a low-energy limit of a more complete theory,
can also lead to 
$\kappa_{Z,W} > 1$~\cite{Alonso:2016btr,Liu:2016idz}.}, where the presence of a doubly-charged scalar can restore unitarity for $\kappa_{V, \text{on}} > 1$. Generalised Georgi-Machacek models, among other attractive features (see e.g.\ Refs.~\cite{Chen:2023bqr,Chen:2025vtg} for recent studies), additionally allow a simultaneous equal enhancement of Higgs couplings to vector bosons and fermions~\cite{Logan:2015xpa}, and the maximum value of $\kappa_{V}$ could be as large as $2.36$ depending on the model and the mass of the doubly-charged scalar~\cite{Logan:2015xpa,OPAL:2002ifx}. 
In such scenarios, however, the off-shell region would be similarly enhanced, while 
the current bound from the off-shell signal strength is 
actually more constraining. Additional effects would therefore be needed to compensate for the on-shell Higgs coupling enhancements in order to make the cross-section rate in the off-shell region compatible with the experimental bounds.

In the following we will focus our analysis on effects that give rise to unitarity cancellations and a reduction of the off-shell contributions,
while we do not explicitly explore the possible sources giving rise to an enhancement of $\kappa_{\text{on}}$.
For this purpose we investigate several types of scenarios in which 
we analyse how much the bound on the total width 
can get
weakened if the assumption of equal on-shell and off-shell coupling modifiers is relaxed, taking into account the latest experimental bounds from searches for BSM particles.
Specifically, 
we concentrate on the case of additional scalars and investigate possible interference effects with the Higgs and $gg \rightarrow ZZ$ continuum that could lead to 
a reduction of the signal strength
in the off-shell region. In particular, in Sect.~\ref{sec:res} we discuss how a propagating scalar in the $s$-channel could 
affect
the Higgs-width 
determination, while in Sect.~\ref{sec:loop} we explore the impact of a coloured scalar in the gluon-fusion loop and a neutral singlet in the Higgs propagator. 
We 
discuss our results and conclude in Sect.~\ref{sec:conc}.

\section{Impact of 
additional scalar propagator contributions on the Higgs width determination}
\label{sec:res}
The $gg \rightarrow H \rightarrow ZZ$ channel is characterised by an increased cross section rate
in the off-shell region, close to the $tt$ mass threshold, and by considerable interference effects with the continuum $gg \rightarrow ZZ$ for large virtuality~\cite{Kauer:2012hd,Kauer:2013cga}. The rate and shape of the $m_{ZZ}$ distribution can be substantially altered 
by contributions
of $s$-channel diagrams 
arising from the propagators of additional scalars
which induce resonant contributions but also interfere with the continuum and with the contribution of the SM-like Higgs
boson~\cite{Maina:2015ela,Kauer:2015hia}. 
A potentially 
destructive interference contribution from new physics decreasing the overall rate in the off-shell regime could 
have the consequence that the determination of the Higgs width based on the assumption of equal on-shell and off-shell coupling modifiers results in a too small value for
the total width of the Higgs boson~\cite{Logan:2014ppa}. 

In order to quantify how sizeable the impact 
of this kind of contributions
on the determination of the Higgs width can be, we investigate this effect in detail including Higgs and continuum interference effects. Assuming the presence of one additional scalar singlet $S$ coupled to the top-quark and the $Z$ boson, the SM Lagrangian can be extended with the effective interactions for the $S$ field as
\begin{equation}
		{\cal{L}} \supset - C_{Stt} \frac{y_t}{\sqrt{2}} \bar{t} S t 
					+ C_{SZZ} \frac{e^2 v}{4 c_W^2 s_W^2} Z_\mu Z^\mu S \;,
\end{equation}
where $y_t$ is the Higgs Yukawa coupling to the top quark in the SM and $e$, $v$ are the electric charge and Higgs vacuum expectation value, respectively. We denote the cosine (sine) of the 
weak mixing angle as $c_W$ ($s_W$). We additionally introduce 
coupling modifiers $\kappa_i$ for the couplings of the SM-like Higgs boson $H$ to the field $i \in \{t, Z\}$. 
In this way the couplings of $H$
to SM particles can be enhanced which would 
give rise to
an increase in the total width.
In the following discussion we do not distinguish between on-shell and off-shell values for 
the ``Lagrangian'' coupling modifiers
$\kappa_Z$ and $\kappa_t$, i.e.\ we consider the case where the reduction of the off-shell contributions results from interference effects rather than from reduced off-shell coupling modifiers of the detected Higgs boson. In the experimental analyses making use of the expressions for the on-shell and off-shell signal strengths as given in \refeq{eq:muon} and \refeq{eq:muoff}, respectively, destructive interference contributions would enter as ``effective'' coupling modifiers $\kappa^{\text{eff}}_{t, \text{off}}$ and 
$\kappa^{\text{eff}}_{Z, \text{off}}$ with 
$\kappa^{\text{eff}}_{t, \text{off}} 
\kappa^{\text{eff}}_{Z, \text{off}} < 
\kappa_{t, \text{on}} \kappa_{Z, \text{on}}$.

Assuming that no additional new physics states affect the unitarisation of the $t\bar{t} \rightarrow ZZ$ scattering apart from $S$, the sum-rule
\begin{equation}
	\label{eq:sumrule}
		\kappa_t \kappa_Z + C_{Stt} C_{SZZ} = 1  
\end{equation}
has to be fulfilled~\cite{Logan:2014ppa}, which relates the 
(Lagrangian)
coupling modifiers of $H$ to the 
BSM couplings
of the additional scalar $S$.

\begin{figure}
	\begin{center}
		\includegraphics[width=0.48\textwidth]{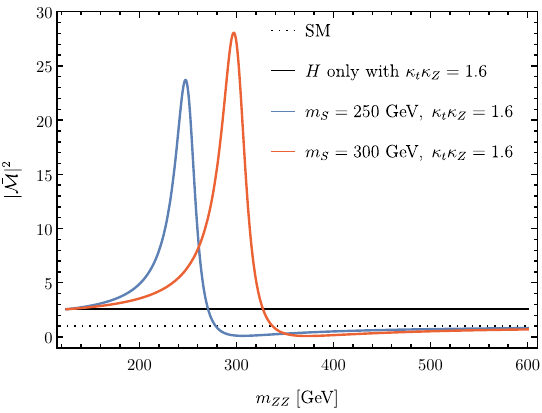}
		\caption{Absolute value squared of the 
        matrix element factor
        expressing the modification of $gg \rightarrow ZZ$ 
        caused by the presence of
        an additional scalar $S$. The black 
        dotted and solid
        lines show the SM value and an enhancement from $\kappa_t \kappa_Z = 1.6$, respectively. The coloured lines show the 
        combined effects of the enhanced coupling modifiers together with the contribution of the
        additional scalar, $gg \rightarrow S \rightarrow ZZ$, 
        for two mass values of $S$. \label{fig:analytical}}
	\end{center}
\end{figure}

The behaviour of the $g g \rightarrow Z Z$ scattering in the off-shell region has been studied analytically in Ref.~\cite{Logan:2014ppa} using the matrix element factor
\begin{equation}
	\bar{{\cal{M}}} = \frac{{\cal{M}}_H + {\cal{M}}_S}{{\cal{M}}_H^\text{SM}} 
					= \kappa_t \kappa_Z - (\kappa_t \kappa_Z - 1) \frac{p^2 - m_H^2}{p^2 - m_S^2 + i m_S \Gamma_S}\;,
\end{equation}
which modifies the matrix element. For demonstration purposes, we plot the absolute value squared of $\bar{{\cal{M}}}$ in Fig.~\ref{fig:analytical}. Expectedly, the 
contribution
of the field $S$ 
leads
to a resonant peak, 
while above the peak, i.e.\ for $m_{ZZ} \gtrsim m_S$, destructive interference contributions give rise to a reduction below the SM value.
For high $m_{ZZ} \gg m_S$, $\bar{\cal{M}}$ 
approaches
the SM value. This destructive interference effect could lead to a reduced event count in the off-shell region. The cut imposed 
on $m_{ZZ}$ in the experimental analyses (e.g.\ the ATLAS analysis of \citere{ATLAS:2023dnm} starts at $220$~GeV) could also potentially remove the resonant peak structure. 
The resulting suppression of the off-shell contribution in combination
with the on-shell signal strengths 
could have the effect that
the indirect Higgs width 
determination assuming equal 
(effective)
on-shell and off-shell coupling modifiers results in a smaller value for
$\GamRat$ than the one that is actually realised in nature.
For comparison we also
show the amplitude squared for the case of $\kappa_t \kappa_Z \sim \sqrt{\mu_\text{off}} \sim 1.6$, i.e.\ the maximum allowed value for the coupling modifiers under the naive interpretation $\kappa_t \kappa_Z = \kappa_{t,\text{off}}^{\text{eff}} \kappa_{Z, \text{off}}^{\text{eff}} \sim \sqrt{\mu_\text{off}}$ and using the 
limit on the off-shell signal strength from
Eq.~\eqref{eq:atlaslim}. Of course, a proper re-interpretation should take into account the effects arising from the new scalar state, as we will further discuss in the following. 

We proceed to a detailed numerical investigation 
with a Monte-Carlo simulation  using {\textsc{FeynRules}}~\cite{Christensen:2008py,Alloul:2013bka} and {\textsc{NLOCT}}~\cite{Degrande:2014vpa} to extract a loop-ready {\textsc{UFO}}~\cite{Degrande:2011ua,Darme:2023jdn} model that can be used with {\textsc{Madgraph5\_MC@NLO}}~\cite{Alwall:2014hca,Hirschi:2015iia}.\footnote{We also use \textsc{FeynArts}, \textsc{FormCalc}, \textsc{LoopTools}~\cite{Hahn:1998yk,Hahn:2000jm,Hahn:2000kx,Hahn:2016ebn} and \textsc{PackageX}~\cite{Patel:2016fam}  for various numerical and analytical checks, throughout this work.} We include one-loop diagrams with top quarks in the loop that contribute to the process $gg \rightarrow ZZ$ with two undecayed on-shell $Z$ bosons.
The generation level cut $\sqrt{\hat{s}} > 220$~GeV is imposed in order to restrict the phase space to the off-shell regime, and $m_{ZZ}$ is defined as the invariant mass of the sum of the two $Z$ boson four-momenta.

It is convenient to parameterise the (differential) cross sections for each scalar mass value $m_S$ as
\begin{flalign}
	\label{eq:xsecpar}
		d\sigma_{gg \rightarrow ZZ}&(\kappa_t \kappa_Z, C_{Stt} C_{SZZ}) = &&\\\nonumber &d\sigma^{}_{(0,0)} + \kappa_t \kappa_Z d\sigma_{(2,0)} + \kappa_t^2 \kappa_Z^2 d\sigma_{(4,0)} &&\\\nonumber &+\; C_{Stt} C_{SZZ} d\sigma_{(0,2)} + \kappa_t \kappa_Z C_{Stt} C_{SZZ} d\sigma_{(2,2)} &&\\\nonumber &+\; C_{Stt}^2 C_{SZZ}^2 d\sigma_{(0,4)}\;,
\end{flalign}
and compute the contributions $d\sigma_{(i,j)} / d m_{ZZ} $, where $i$ ($j$) stands for the Higgs coupling (BSM scalar coupling) order, which arise from different diagrams (see Tab.~\ref{tab:contrs}). These contributions are computed for fixed values $\kappa_t = \kappa_Z = 1$ and $C_{Stt} = C_{SZZ} = 1$ and are used to construct the full cross section $\sigma_{gg \rightarrow ZZ}$. The contributions with non-zero BSM scalar coupling dependence have an additional dependence on the mass of the scalar $m_S$ which is kept implicit. The initial gluons are extracted from the colliding protons using parton distribution functions (we use the \verb|nn23nlo| PDF). We do not include contributions that arise from quarks in the initial state of the hard process.

\begin{figure}
	\begin{center}
		\includegraphics[width=0.48\textwidth]{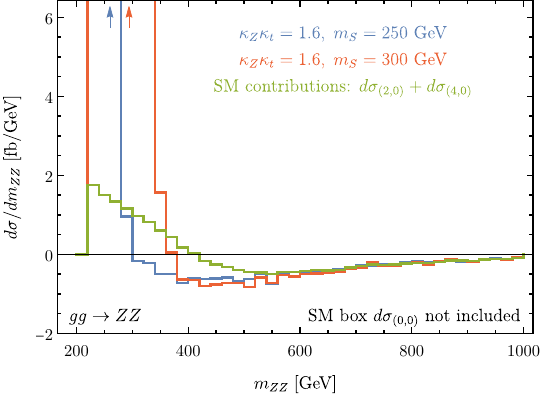}
		\caption{Histograms for $gg \rightarrow H,S \rightarrow ZZ$ for two values of $m_S$, without including contributions from the square of the SM box diagrams (the interference contributions involving the box diagrams are included). A peak is present where $S$ goes on-shell at $m_{ZZ} = m_S$, 
        while for higher values of $m_{ZZ}$
        a decrease of events occurs as
        compared to the SM bin heights. 
        \label{fig:histo}}
	\end{center}
\end{figure}

We show the differential cross section $d \sigma / d m_{ZZ}$ in Fig.~\ref{fig:histo}. We 
leave out
the contribution from $d\sigma_{(0,0)}$ which corresponds to the pure squared box diagrams $g g \rightarrow ZZ$ but include all other interferences, in order to showcase the presence of 
destructive interference contributions for larger values of $m_{ZZ}$, in agreement with the analytical considerations of Fig.~\ref{fig:analytical}. This is shown for 
$\kappa_Z \kappa_t = 1.6$ and two mass values for $S$, 
$m_S = 250$~GeV and $m_S = 300$~GeV.

\begin{table}
	\begin{center}
	\begin{tabular}{lccc}
		\toprule
		$d \sigma_{(i,j)}$ & $\kappa_t = \kappa_Z$ & $C_{Stt} = C_{SZZ}$ & Contribution \\
		\midrule
		$d \sigma_{(0,0)}$ & 0						 & 0					& SM box  \\
		$d \sigma_{(2,0)}$ & 1						 & 0					& Higgs-box interference \\
		$d \sigma_{(4,0)}$ & 1						 & 0					& Higgs propagator \\
		$d \sigma_{(0,2)}$ & 0						 & 1					& $S$-box interference \\
		$d \sigma_{(2,2)}$ & 1						 & 1					& $S$-Higgs interference \\
		$d \sigma_{(0,4)}$ & 0						 & 1					& $S$ resonance \\
	\end{tabular}
		\caption{
        The relevant contributions for the employed parameterisation of differential cross sections 
        using the specified fixed values for the Higgs coupling modifiers $\kappa_t$, $\kappa_Z$ and the $S$ couplings $C_{Stt}$, $C_{SZZ}$. 
        \label{tab:contrs}}
	\end{center}
\end{table}

\begin{figure*}[t!]
	\includegraphics[width=0.49\textwidth]{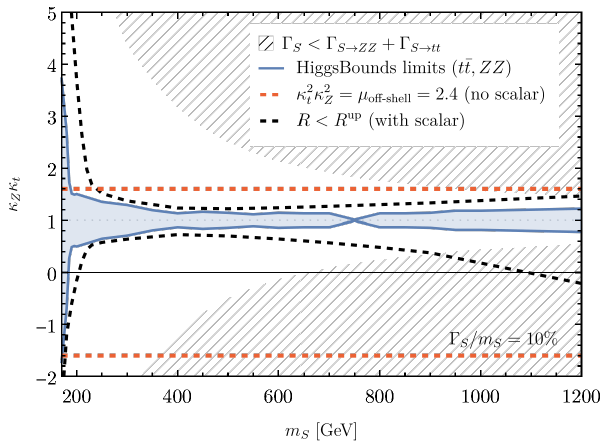}
	\hfill
	\includegraphics[width=0.49\textwidth]{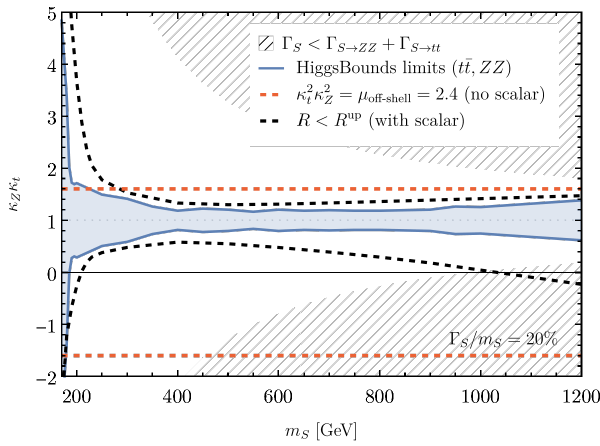}
	\caption{The allowed regions for the product of coupling 
    modifiers $\kappa_t \kappa_Z$ from the off-shell Higgs contributions to $gg \to ZZ$ are indicated by the black dashed contours for the case where an additional scalar $S$ is present with
    $\Gamma_S / m_S = 10\%$ ($\Gamma_S / m_S = 20\%$) on the left (right).
    In the region between the two black dashed contours the bound
    $R(\kappa_t \kappa_Z, C_{Stt} C_{SZZ}) < R^\text{up}$ 
    with the imposed sum-rule $\kappa_t \kappa_Z + C_{Stt} C_{SZZ} = 1$ holds, while the hatched region is theoretically inaccessible for the assumed value of $\Gamma_S / m_S$.
    The red dashed line shows the limit 
    of $\kappa_t^2 \kappa_Z^2 = 2.4$ that would be obtained for the case where the additional scalar is absent. 
    The \textsc{HiggsBounds} limits from searches for additional scalars are indicated by the blue contours, 
    and the blue area corresponds to the region that is allowed by these limits. 
 \label{fig:kappalims}}
\end{figure*}

As mentioned above,
interpreting the off-shell signal strength limit of Eq.~\eqref{eq:atlaslim} in terms of the coupling 
modifiers, $\mu_\text{off-shell} = \kappa_t^2 \kappa_Z^2$, 
yields the naive
limit of $\kappa_t \kappa_Z \lesssim 1.55$ on the coupling modifiers. However, the interference contributions enter in terms of the square root of the signal strength $\sqrt{\mu_\text{off-shell}}$, which are also included in the analysis by ATLAS~\cite{ATLAS:2023dnm}. Thus, in order to properly include the interference effects and quantify the impact on the $\kappa_t \kappa_Z$ limits from the additional scalar, we define the ratio between BSM and SM cross sections as
\begin{equation}
	R(\kappa_t \kappa_Z , C_{Stt} C_{SZZ}) = \frac{\sigma_{gg\rightarrow ZZ}(\kappa_t \kappa_Z, C_{Stt} C_{SZZ})}{\sigma_{gg \rightarrow ZZ}^\text{SM}}\;,
    \label{eq:Rratio}
\end{equation}
where $\sigma_{gg \rightarrow ZZ}^\text{SM}$ can be obtained from Eq.~\eqref{eq:xsecpar} 
if
no BSM contributions are included and the coupling modifiers are set to unity, i.e.~$\sigma_{gg \rightarrow ZZ}(1, 0)$. The maximum value of $\mu_\text{off}^\text{up} < 2.4$ can be subsequently reinterpreted as an upper limit on 
$R$ given by
\begin{equation}
	R^\text{up} = \frac{\sigma_{(0,0)} + \sqrt{\mu_\text{off}^\text{up}} \, \sigma_{(2,0)} + \mu_\text{off}^\text{up} \, \sigma_{(4,0)}}{\sigma_{(0,0)} +  \sigma_{(2,0)} +  \sigma_{(4,0)}} \simeq 3.4\;.
\end{equation}
Enforcing the sum rule~\eqref{eq:sumrule} allows us to identify parameter values of $\kappa_Z \kappa_t$ that are not excluded by the off-shell signal strength measurement for different values of $m_S$ through the relation $R(\kappa_Z \kappa_t, 1 - \kappa_Z \kappa_t) < R^\text{up}$. 
We perform a scan calculating the ratio for different values of $m_S$ and show the limit from this ratio in \reffi{fig:kappalims} by the black dashed contours. 
For the majority of different scalar mass values, using the limit $\mu_\text{off-shell} = \kappa_t^2 \kappa_Z^2 < 2.4$ within the BSM scenario would be conservative as the additional events from the 
resonant contribution of the additional scalar
would restrict the allowed values of $\kappa_t \kappa_Z$ to smaller values. 
In contrast, the 
reduction of the cross section
caused by the interference for small mass values of $S$ would allow larger values of $\kappa_t \kappa_Z$. 
In the on-shell analyses of the detected Higgs boson at 125~GeV the enhanced values of those coupling modifiers could be compensated by an increase in the total width of the Higgs boson.

It is important to note in this context that 
bounds on $C_{Stt} C_{SZZ}$ from searches for additional scalars 
imply a bound on $\kappa_t \kappa_Z$ through the sum-rule of Eq.~\eqref{eq:sumrule}. 
In order to investigate 
how the existing limits from searches for additional scalars affect the allowed
values of $\kappa_t \kappa_Z$ 
in the considered mass range of an additional scalar particle,
we utilise {\textsc{HiggsBounds-5}}~\cite{Bechtle:2020pkv,Bahl:2021yhk}\footnote{We have modified \textsc{HiggsBounds} 
to include the results of Ref.~\cite{CMS-PAS-HIG-24-002} 
which yields slightly improved bounds.} through the {\textsc{HiggsTools}} framework~\cite{Bahl:2022igd}.
The decays of $S$ are model-dependent and could open different channels for the 
detection of 
this new particle. As such, any bound on $C_{Stt}$ and $C_{SZZ}$ is inherently model-dependent. For simplicity we assume that $S$ decays to top quarks and $Z$ bosons with the same partial widths as the detected Higgs boson at 125~GeV up to the factors $C_{Stt}$ and $C_{SZZ}$, respectively. We allow the possibility of additional undetectable decays that contribute to the total width of $S$, which is 
treated
as a free parameter. We fix $C_{Stt} = 1$ and determine the allowed values of $C_{SZZ}$ for different masses $m_S$ at fixed total widths. The results after taking into account the 
sum-rule of Eq~\eqref{eq:sumrule} are shown by the blue contours in \reffi{fig:kappalims}. The  allowed area that is compatible with  the experimental  limits from searches for additional scalars is shown in blue. While the experimental results for the off-shell signal strength by themselves would allow significantly larger values of $\kappa_t \kappa_Z$ for the case where interference contributions with an additional scalar $S$ are present compared to the case where no such interference contributions occur, the existing limits from searches for additional scalars put severe constraints on such a scenario. In particular, in our investigation we find that $\kappa_t^2 \kappa_Z^2 > 2.4$ occurs only in a small mass region of $m_S$ close to the $ZZ$ threshold (as discussed above, the detailed impact of the existing search limits is affected by a certain model dependence).

\begin{figure}
	\begin{center}
		\includegraphics[width=0.48\textwidth]{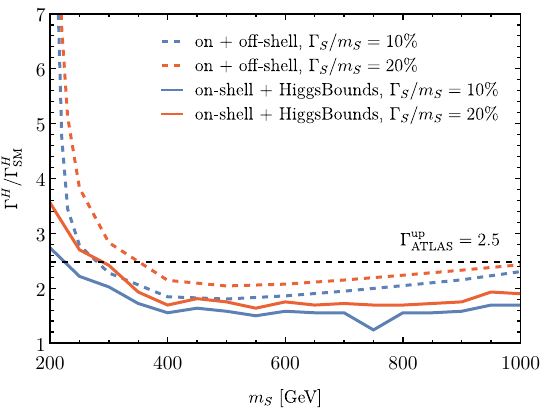}
		\caption{Upper limits on the Higgs width ratio 
        $\Gamma^H / \Gamma^H_\text{SM}$ 
        as a function of $m_S$
        for the case where an additional propagating scalar $S$ contributes in the off-shell region (but is assumed not to
        affect the on-shell region).
        The upper limits that are obtained
        from the combination of the on-shell and off-shell results
        are indicated by the dashed lines
        for two values of $\Gamma_S/m_S$.
        The upper bounds on $\Gamma^H / \Gamma^H_\text{SM}$ arising 
        from the limits from direct Higgs searches 
        implemented via \textsc{HiggsBounds} are shown by the solid lines,
        where the on-shell signal strength is used 
        to infer an upper bound on the total width from the bound on the coupling modifiers.
        \label{fig:gammalims}}
	\end{center}
\end{figure}

In order to quantify the impact of the discussed effects on the determination of the total width of the detected Higgs boson,
we now combine the obtained
upper limit on $\kappa_t \kappa_Z$ 
for different values of the mass of an additional scalar
from the experimental results for the off-shell signal strength
with the on-shell measurements.
We assume that there is no impact from the additional scalar on the on-shell region, which should be a reasonable assumption as long as there is sufficient separation between the 
mass of the detected Higgs boson of 125~GeV and $m_S$.
To demonstrate the effect on the total Higgs width we consider the largest possible value for $\kappa_Z \kappa_t$ from our off-shell results 
that 
is compatible with
the condition $R(\kappa_Z \kappa_t, 1 - \kappa_Z \kappa_t) < R^\text{up}$. 
As before, the on-shell signal strength for this scenario is expressed 
via $\mu_\text{on} = \kappa_t^2 \kappa_Z^2 / (\GamRat)$. 
Utilising the ATLAS on-shell measurement $\mu_\text{on} = 1.01^{+0.23}_{-0.20}$~\cite{ATLAS:2020rej} 
allows us to obtain the range of $\GamRat$ that is compatible with the on-shell result from ATLAS for a particular mass 
value $m_S$. This is shown in \reffi{fig:gammalims} 
for two values of $\Gamma_S/m_S$.
Furthermore, 
we display the bounds on the total Higgs width that can be inferred from 
the limits from direct Higgs searches, implemented via \textsc{HiggsBounds}, 
where also the on-shell signal strength
is used to relate the upper bound on the coupling modifiers to an upper bound on the total width.
We find that the latter bounds are
stronger than the ones obtained from the on-shell and off-shell signal strengths.
For large values of the mass of the additional scalar
the upper 
limits on $\GamRat$ 
from the on-shell and off-shell signal strengths
get close to the ATLAS limit of $2.5$. 
For intermediate values of $m_S \gtrsim 300$~GeV the upper limit on $\GamRat$ is lower due to the additional events from the resonant scalar (with couplings related to $\kappa_t \kappa_Z$ through the sum-rule), 
and the 
application
of the experimental limit of $2.5$ is 
conservative in such BSM models. 
As expected from the discussion above, the region where the actual value of $\GamRat$ can exceed the value that is obtained from the experimental analyses assuming equal on-shell and off-shell coupling modifiers is restricted to the low-mass region
$m_S \lesssim 300$~GeV 
of the additional scalar and is severely constrained by the existing Higgs search limits. We find that for the investigated scenarios an increase of $\GamRat$ of at most {40\%} compared to the indirect determination relying on theoretical assumptions is possible for the smallest values of $m_S$. There is potential for further tightening this bound because resonant effects of the additional scalar could be detected in the control regions of the experimental analyses~\cite{ATLAS:2023dnm}.

\section{Impact 
of BSM scalar loop effects on the Higgs width determination}
\label{sec:loop}

While above we have analysed the tree-level propagator-type contribution of an additional scalar, we will now investigate loop-level effects of BSM particles that interact with the detected Higgs boson.
The sensitivity of off-shell contributions to $gg \rightarrow ZZ$ for probing
new physics through potentially sizeable loop effects 
has previously been investigated in \citeres{Englert:2014ffa,Goncalves:2017iub,Lee:2018fxj}.

\subsection{Impact 
of
a coloured scalar in the gluon-fusion loop}

\begin{figure}[t!]
\subfloat{\includegraphics[width=0.245\textwidth]{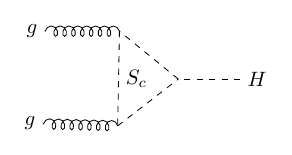}}
	\hspace{-0.4cm}
\subfloat{\includegraphics[width=0.245\textwidth]{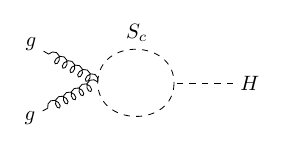}}
	\caption{Additional one-loop diagrams contributing to gluon-fusion production that arise from the presence of an additional coloured scalar.
 \label{fig:feyndiags}}
\end{figure}

\begin{figure*}[t]
\includegraphics[width=0.49\textwidth]{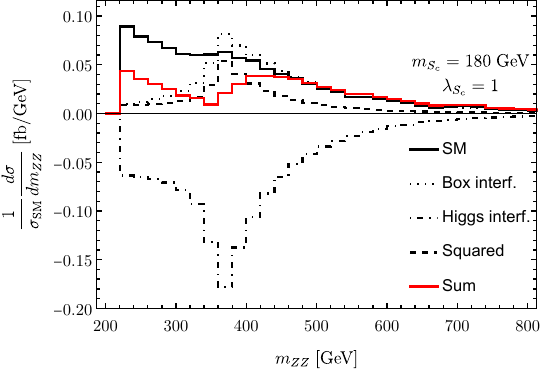}
\hfill
\includegraphics[width=0.49\textwidth]{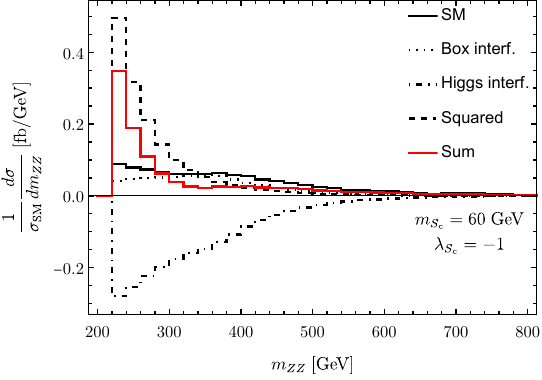}
	\caption{Example histograms for the $gg \rightarrow ZZ$ process for 
    $m_{S_c} = 180$~GeV, $\lambda_{S_c} = 1$ (left) and 
    $m_{S_c} = 60$~GeV, $\lambda_{S_c} = -1$ (right). The different interference effects of the BSM contribution with the box and Higgs contributions are shown together with the sum of all contributions. The BSM contribution squared is also 
    indicated.
 \label{fig:colscalhisto}}
\end{figure*}

In order to investigate the effects of the presence of an additional field contributing to the gluon fusion process at one-loop level, we extend the SM with a coloured complex scalar $S_c$ that is a singlet under the $SU(2)$ and a triplet under the $SU(3)$.  The Lagrangian reads
\begin{align}
    {\cal{L}}  \supset D_\mu S_c D^\mu \bar{S}_c - m_{S_c}^2 S_c \bar{S}_c
	+ \lambda_{S_c} \Phi^\dagger \Phi S_c \bar{S}_c \;,
\end{align}
where $m_{S_c}$ denotes the mass of the scalar and $\lambda_{S_c}$ the coupling to the Higgs field $\Phi = \left(0, (v+H)/\sqrt{2}\right)$ (in unitary gauge). The field $S_c$ couples to the gluons through the covariant derivative and to the physical Higgs field through the interaction $\lambda_{S_c} v H  S_c \bar{S}_c$ 
inducing the additional Feynman diagrams contributing to the production of the Higgs boson shown in Fig.~\ref{fig:feyndiags}. As before, we allow further modifications of the Higgs coupling though $\kappa$ modifiers and utilise the same toolchain as in the previous section.

\begin{table}[t]
	\begin{center}
	\begin{tabular}{lcccc}
		\toprule
		$d \sigma_{(i,j)}$ & $\kappa_t$ & $\kappa_Z$ & $\lambda_{S_c}$ & Contribution \\
		\midrule
		$d \sigma_{(0,0)}$ & 0		& 0				 & 0					& SM box  \\
		$d \sigma_{(2,0)}$ & 1		& 1				 & 0					& Higgs ($t$ loop) -- box interference \\
		$d \sigma_{(4,0)}$ & 1		& 1				 & 0					& Higgs ($t$ loop) propagator \\
		$d \sigma_{(1,1)}$ & 0		& 1				 & 1					& Higgs ($S_c$ loop) -- box interference \\
		$d \sigma_{(3,1)}$ & 1		& 1				 & 1					& Higgs ($S_c$ loop) -- ($t$ loop) interference \\
		$d \sigma_{(2,2)}$ & 0		& 1				 & 1					& Higgs ($S_c$ loop) propagator \\
	\end{tabular}
		\caption{
        The contributions for the differential cross section parameterisation in the presence of a coloured scalar $S_c$ 
        in the loop
        using the specified fixed values for $\kappa_t$, $\kappa_Z$ and $\lambda_{S_c}$. 
        \label{tab:contrs_Sc}}
	\end{center}
\end{table}

The parameterisation of the cross section in this case 
is given by
\begin{flalign}
	\label{eq:xsecpar_colscal}
		d\sigma_{gg \rightarrow ZZ}&(\kappa_t, \kappa_Z, \lambda_{S_c}) = &&\\\nonumber &d\sigma^{}_{(0,0)} + \kappa_t \kappa_Z d\sigma_{(2,0)} + \kappa_t^2 \kappa_Z^2 d\sigma_{(4,0)} &&\\\nonumber 
		&+\; \kappa_Z \lambda_{S_c} d\sigma_{(1,1)} + \kappa_t \kappa_Z^2 \lambda_{S_c} d\sigma_{(3,1)} &&\\\nonumber &+\;  \kappa_Z^2 \lambda_{S_c}^2 d\sigma_{(2,2)}\;.
\end{flalign}
For this scenario, the first number in our $(i,j)$ notation stands for the order of the Higgs $\kappa$-modifier coupling, while the second is the order of the $S_c$--Higgs coupling. Thus, $d\sigma_{(1,1)}$ 
corresponds to the interference of the diagrams in Fig.~\ref{fig:feyndiags} (with $H \rightarrow ZZ$) with the box diagrams of $gg \rightarrow ZZ$. Similarly, $d\sigma_{(3,1)}$
denotes
the interference of the BSM diagrams with the SM-type gluon fusion, and $\sigma_{(2,2)}$ 
stands for the BSM production squared (see also Tab.~\ref{tab:contrs_Sc}). We note that the first line is the same as Eq.~\eqref{eq:xsecpar} 
as it corresponds to the pure 
SM-like contribution. 
The BSM interference with the Higgs contributions, $d\sigma_{(3,1)}$, is particularly important in this scenario as it can be larger than the interference with the continuum, $d\sigma_{(1,1)}$. It can be either constructive or destructive, as its behaviour flips depending on the sign of $\lambda_{S_c}$.
Modifications of the Higgs $\kappa$ factors could further enhance the $d \sigma_{(3,1)}$ contribution
which scales as $\kappa^3$. 
Example histograms are shown in Fig.~\ref{fig:colscalhisto}.

\begin{figure}[h!]
\includegraphics[width=0.45\textwidth]{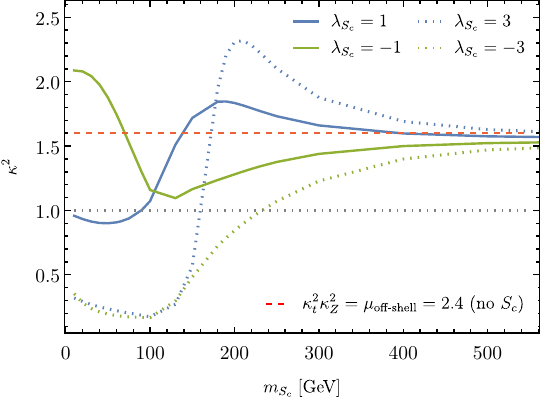}\\
	\caption{Upper limits on the Higgs coupling modifier $\kappa = \kappa_t = \kappa_Z$ squared for the case where a coloured scalar contributes to the gluon-fusion production for different values of $\lambda_{S_c}$. The red dashed line shows the $\kappa_t^2 \kappa_Z^2 = \mu_\text{off}$ interpretation without the additional state.
 \label{fig:colscallims}}
\end{figure}

The ratio $R(\kappa_t, \kappa_Z, \lambda_{S_c})$ is constructed as the BSM cross section $d\sigma_{gg \rightarrow ZZ} (\kappa_t, \kappa_Z, \lambda_{S_c})$ 
divided by the SM one, similarly to \refeq{eq:Rratio}. Unlike the previous case the cross section (and thus the ratio) follows a different scaling with respect to the Higgs coupling modifiers. Perturbative unitarity in this case is more involved and the condition of \refeq{eq:sumrule} is not applicable in this case. Hence, the parameter $\lambda_{S_c}$ is kept free, and assuming $\kappa = \kappa_t = \kappa_Z$ we show the impact on $\kappa^2$ in Fig.~\ref{fig:colscallims} by imposing $R(\kappa, \kappa, \lambda_{S_c}) < R^\text{up}$. 
The results are shown as a function
of $m_{S_c}$ for different values of $\lambda_{S_c}$.
The sign of $\lambda_{S_c}$ is of particular importance as positive values induce 
destructive interference effects for masses 
$m_{S_c} \gtrsim 125$~GeV (indicated by the weaker bound on $\kappa^2$), while 
for the case $\lambda_{S_c} = -1$ destructive interference occurs for smaller masses. The 
absolute value of the parameter $\lambda_{S_c}$ affects how sizeable the interference contributions are. In all cases the limit 
approaches $1.6$ 
for large values of the mass $m_{S_c}$ as expected.

\begin{figure}[h!]
\includegraphics[width=0.45\textwidth]{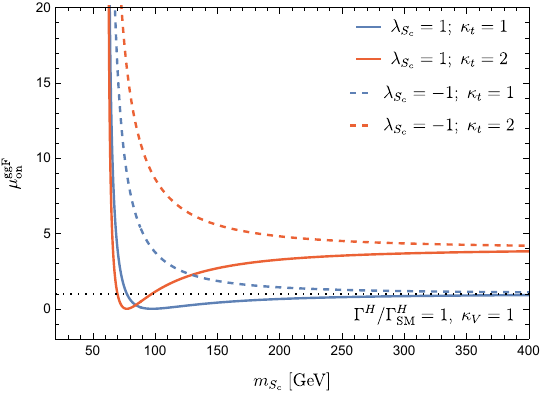}\\
	\caption{
    On-shell signal strength $\mu^\text{ggF}_\text{on}$ 
    as function of the mass $m_{S_c}$ of the coloured scalar in the gluon fusion loop for the case where $\Gamma^H / \Gamma^H_\text{SM} = 1$ and $\kappa_V = 1$ are fixed.
    Different combinations of $\lambda_{S_c}$ and $\kappa_t$ are shown. 
 \label{fig:colscal_muon} 
	}
\end{figure}

The loop contribution of the additional scalar makes the interplay between the off-shell and the on-shell signal strengths more involved than for the case of a lowest-order contribution discussed above.
An important constraint on the considered scenario arises from the fact that
the additional scalar loop contributes to the on-shell production of the Higgs boson via gluon-fusion but not to the weak boson fusion production. 
Requiring consistency of the predictions with the experimental results for
both production modes 
limits the possible enhancement of the Higgs-boson width.
The weak boson fusion signal strength in the on-shell region 
is given by
$\mu^\text{WBF}_\text{on} = \kappa_V^4 / (\Gamma^H / \Gamma^H_\text{SM})$,
where for simplicity we have assumed
$\kappa_Z = \kappa_W = \kappa_V$. We estimate the impact on the on-shell gluon fusion signal strength as
\begin{equation}
	\label{eq:colscal_muggf}
	\mu^\text{ggF}_\text{on} = \frac{\kappa_V^2}{\Gamma^H / \Gamma^H_\text{SM}} \left\lvert \kappa_t + 
    \frac{\lambda_{S_c} v^2 \left[ \tau_{S_c} f(\tau_{S_c}) -1 \right]}{m_H^2 \tau_t \left[ (\tau_t - 1) f(\tau_t) - 1\right]} 
    \right\rvert^2\;,
\end{equation}
where $\tau_i = 4 m_i^2 / m_H^2$ for $i \in \{t, S_c\}$.\footnote{For the form of the loop contributions from particles of different spin, see e.g.\ Ref.~\cite{Djouadi:2005gj}.} The function $f(\tau_i)$ describes the triangle loop of the top quark or coloured scalar and is given by~\cite{Spira:1995rr,Djouadi:2005gi} (we use the conventions of~\citere{Plehn:2009nd})
\begin{equation}
	\label{eq:loopf}
	f(\tau_i) = \begin{cases}
					\arcsin^2 \tau_i^{-1/2} & \tau_i > 1 \\
					-\frac{1}{4} \left[ \log{\frac{1+\sqrt{1-\tau_i}}{1 - \sqrt{1-\tau_i}}} - i \pi \right]^2 & \tau_i < 1\;.
				\end{cases}
\end{equation}
Fixing in a first step the Higgs width ratio to the SM value and $\kappa_V$ to unity,
the behaviour of the gluon-fusion signal strength $\mu^\text{ggF}_\text{on}$ is shown in Fig.~\ref{fig:colscal_muon}.
For low values of the coloured scalar mass $m_{S_c}$ 
the signal strength $\mu^\text{ggF}_\text{on}$
increases drastically, which 
could be compensated by an enhanced value of
$\Gamma^H / \Gamma^H_\text{SM}$.
As shown in the figure, for $\kappa_t = 2$ an increased value of $\mu^\text{ggF}_\text{on}$ also remains for large values of $m_{S_c}$ where the loop contribution of the coloured scalar becomes small.

\begin{figure}[h!]
\includegraphics[width=0.45\textwidth]{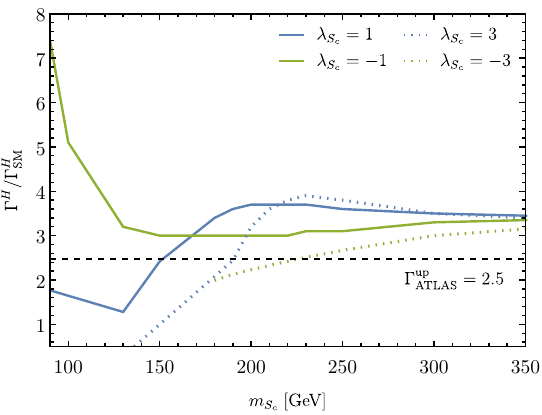}\\
	\caption{
		Upper limits on $\GamRat$ 
        as function of the mass $m_{S_c}$ of the coloured scalar in the gluon fusion loop
        for different values of $\lambda_{S_c}$. 
        The upper limits arise from 
        the $95\%$~CL bounds in the plane of the on-shell signal strengths $\mu_\text{on}^\text{ggF}$ and $\mu_\text{on}^\text{WBF}$
        together with the bounds on
        the Higgs coupling modifiers $\kappa = \kappa_t = \kappa_V$ obtained from the off-shell results in Fig.~\ref{fig:colscallims}. 
        Smaller values of $m_{S_c}$ 
        are incompatible with the bounds on the on-shell signal strengths
        and are not shown.
 \label{fig:colscalwidthlims}}
\end{figure}

In order to determine how large values of the total Higgs-boson width are allowed in this case,
we use the $95\%$~CL constraints in the $\mu^{ggF}_\text{on}$--$\mu^{WBF}_\text{on}$ plane from \mbox{ATLAS} in the $4\ell$ channel~\cite{ATLAS:2020rej} and require that the predicted on-shell signal strengths incorporating the loop contribution of the coloured scalar do not lie outside of the allowed contour. 
The Higgs coupling modifiers $\kappa = \kappa_t = \kappa_V$ are set to the maximum value allowed by the off-shell results that we obtained in Fig.~\ref{fig:colscallims}, i.e.\ $R(\kappa, \kappa, \lambda_{S_c}) = R^\text{up}$. 
Subsequently we vary the Higgs width ratio $\Gamma^H / \Gamma^H_\text{SM}$ to obtain the maximum value that does not violate the on-shell constraints, which is shown in Fig.~\ref{fig:colscalwidthlims} for different values of $\lambda_{S_c}$ and masses $m_{S_c}$. For $\lambda_{S_c} = -1$ 
and $m_{S_c} \sim 90$~GeV the maximum value reaches $\sim 7.5$, 
while lower values of $m_{S_c}$ would yield predicted values for the signal strengths that are outside of the allowed
on-shell experimental contour. 
For higher values of $\lambda_{S_c}$ the maximum of $\Gamma^H / \Gamma^H_\text{SM}$ within the bounds on the on-shell signal strengths is obtained for heavier masses of the scalar, but the width ratio does not exceed $\sim 4$ in this case.
This is only slightly higher than the 
bound of $\sim 3.5$ that is obtained for the highest values of $m_{S_c}$ where the contribution of the scalar loop is small. 
It should be noted that the reason why a somewhat higher bound is obtained here compared to the value of $\GamRat < 2.5$ quoted in \refeq{eq:atlaslim} is related
to the fact that here the weaker bounds from the two-dimensional contour of the on-shell signal strengths are used instead of the combined on-shell measurement. 

As a result of this analysis we find that a significant enhancement of the total Higgs-boson width above the limit that has been obtained for the case without the additional scalar is only possible for very small values of the mass of the additional scalar. While the detailed prospects for direct searches depend on the specific model under consideration, it is obvious that scenarios with a light coloured scalar particle are tightly constrained by direct searches for coloured resonances. In particular,
decays of the scalar triplet 
would require a non-zero hypercharge, while decays to gluons are only possible if one goes beyond scalar triplets (octets could decay to a gluon pair at loop-level). The coloured triplet could be stable and hadronise\footnote{This is similar to the formation of $R$-hadrons due to squarks in supersymmetric models.} with a fractional electric charge, for which current experimental results would typically imply a bound around 1~TeV~\cite{CMS:2016kce}, much larger than the masses that could significantly affect the Higgs width determination. For a more detailed discussion regarding decays of coloured scalars, see Ref.~\cite{Preuss:2021ebd}.

\subsection{Impact 
of a
scalar loop in the Higgs propagator}

As a further scenario in this context we study
the effects from a scalar field in the $s$-channel Higgs propagator 
(a similar scenario was investigated in Ref.~\cite{Goncalves:2017iub}). A scalar field $S$ could potentially impact only the Higgs propagator but not interactions to other SM fields. This can be realised in Higgs-portal scenarios (e.g.~\citeres{Binoth:1996au,Schabinger:2005ei,Patt:2006fw,Ahlers:2008qc,Batell:2009yf,Englert:2011yb}) with an unbroken ${\mathbb{Z}}_{2}$ symmetry, where $S$ couples to the Higgs field via
\begin{align}
    {\cal{L}}  \supset  - \lambda_S S^2 \Phi^\dagger \Phi \;.
\end{align}

\begin{figure}[t!]
\includegraphics[width=0.48\textwidth]{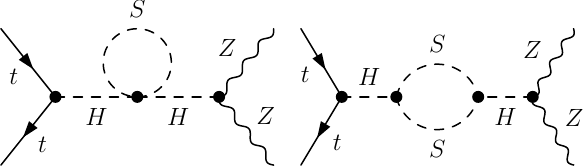}\\
\includegraphics[width=0.48\textwidth]{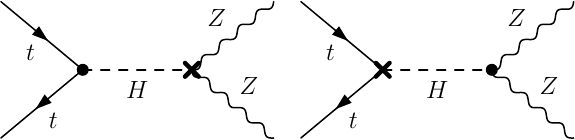}\\
\includegraphics[width=0.24\textwidth]{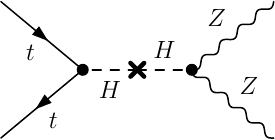}\\
	\caption{
		Feynman diagrams that have to be included to calculate the effect of 
        the loop contribution of
        an additional singlet scalar $S$ to the Higgs propagator. The first line indicates the one-loop contributions to the Higgs self-energy, while the other displayed diagrams contain counterterm contributions.
 \label{fig:feyndiags_portal}}
\end{figure}

\begin{figure}[t!]
\includegraphics[width=0.48\textwidth]{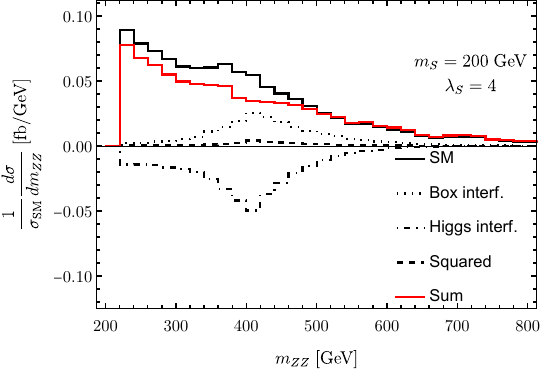}\\
	\caption{
		Example histograms for $gg \rightarrow ZZ$ that show the BSM contribution squared as well as interference contributions with the box and pure-Higgs diagrams and the sum of all contributions in comparison to the SM prediction. 
 \label{fig:hist_portal}}
\end{figure}

After electroweak symmetry breaking the Higgs portal scalar $S$ contributes to $gg \rightarrow H \rightarrow ZZ$ through the diagrams shown in Fig.~\ref{fig:feyndiags_portal}, resulting in the modification factor~\cite{Englert:2020gcp}
\begin{equation}
	\label{eq:loopmod}
	\begin{split}
		1 - \frac{{\cal{M}}^\text{NLO}}{{\cal{M}}^\text{LO}} &=  \frac{\lambda_S^2 v^2}{8 \pi^2 (p^2 - m_H^2)} \\ &\times \big[ B_0(p^2, m_S^2, m_S^2)  - \textrm{Re}B_0(m_H^2, m_S^2, m_S^2) \big]\;,
	\end{split}
\end{equation}
where $m_S$ denotes the mass of the scalar. 
We focus here on the case $m_S > m_H / 2$, so that  
$\textrm{Re}B_0(m_H^2, m_S^2, m_S^2) \equiv B_0(m_H^2, m_S^2, m_S^2)$.
In order to perform a similar study as for
the previous scenarios, we modify the SM {\textsc{UFO}} model file and introduce the BSM factor to simulate signal events. With the same notation as before, the cross section is parameterised as
\begin{flalign}
	\label{eq:xsecpar_portal}
		d\sigma_{gg \rightarrow ZZ}&(\kappa_t, \kappa_Z, \lambda_{S}) = &&\\\nonumber &d\sigma^{}_{(0,0)} + \kappa_t \kappa_Z d\sigma_{(2,0)} + \kappa_t^2 \kappa_Z^2 d\sigma_{(4,0)} &&\\\nonumber 
		&+\; \kappa_t \kappa_Z  \lambda_{S}^2 d\sigma_{(2,2)} + \kappa_t^2 \kappa_Z^2 \lambda_{S}^2  d\sigma_{(4,2)} &&\\\nonumber &+\; \kappa_t^2 \kappa_Z^2 \lambda_{S}^4 d\sigma_{(4,4)}\;,
\end{flalign}
and
the allowed region for $\kappa = \kappa_t = \kappa_Z$ 
is calculated
using the condition
\begin{equation}
	R(\kappa_t, \kappa_Z, \lambda_S) = \frac{\sigma_{gg \rightarrow ZZ} (\kappa_t, \kappa_Z, \lambda_S)}{\sigma_{gg \rightarrow ZZ}^\text{SM}} < R^\text{up}\;.
\end{equation}
As shown in Fig.~\ref{fig:hist_portal}, while the interference with the $gg \rightarrow ZZ$ box diagram is 
constructive, the interference with the leading-order Higgs contribution is 
destructive. This can reduce the rate in the off-shell $ZZ$ region, which 
affects
the Higgs width measurement. 
This is the case despite the fact that the squared BSM contribution by itself is quite small, as indicated by the ``Squared'' 
histogram in the plot.

\begin{figure}[]
\includegraphics[width=0.45\textwidth]{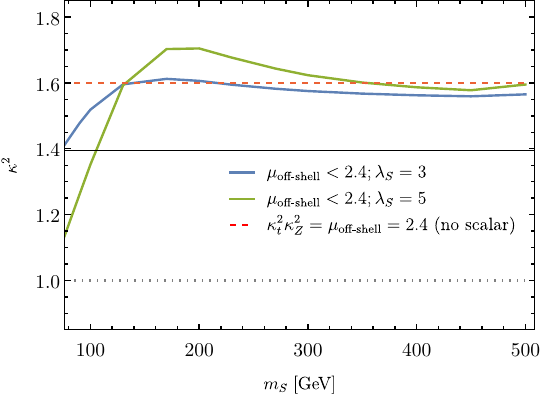}
	\caption{Limits on $\kappa^2 = \kappa_t^2 = \kappa_Z^2$ 
    from off-shell Higgs-boson production for the case where
    an additional scalar enters 
    via a loop contribution to the Higgs-boson propagator.
    The obtained limits are shown as a function of the mass
    of the scalar, $m_S$, for different values of the coupling of the additional scalar to the Higgs boson, $\lambda_S$. 
 \label{fig:portallims}}
\end{figure}

The re-interpretation of the off-shell Higgs signal strength within this model is shown in Fig.~\ref{fig:portallims}. In general the impact on the limits on $\kappa$ is suppressed due to the rather small corrections of Eq.~\eqref{eq:loopmod} (see also \citeres{Englert:2019zmt,Englert:2020gcp}). 
Going beyond the bound of $\kappa^2 \lesssim 1.6$ that would be obtained for the case where the additional scalar in the loop is absent
requires a substantial value of $\lambda_S \sim 5$ for relatively small masses of $m_S \sim 200$~GeV. In principle such large values of $\lambda_S$ could be excluded by the Higgs signal strength measurements at the LHC
if one assumes that the SM is only extended with the field $S$~\cite{Englert:2020gcp}. 

The effect of the modification of Eq.~\eqref{eq:loopmod} for the case where the Higgs goes on-shell, $p^2 = m_H^2$, 
results in a contribution involving the derivative of the $B_0$ function in \refeq{eq:loopmod} with respect to $p^2$ (corresponding to the scalar loop contribution to the Higgs field renormalisation).
The impact on
the on-shell signal strength is that the result given as in the case without the additional scalar, depending only on the Higgs width and $\kappa_t^2 \kappa_Z^2$, is modified by the factor $\lambda_S^2 v^2/(8 \pi^2) B_0^\prime$, 
where $B_0^\prime$ denotes the derivative of the $B_0$ function evaluated at $p^2 = m_H^2$.
For mass values around $m_S = 200\;\text{GeV}$ and $\lambda_S = 5$, 
for which in \reffi{fig:portallims} the limit on $\kappa^2$ is weakest, we find that the shift in the on-shell signal strength amounts to about 10\%. This is compatible with the experimental results at the current level of uncertainties. The effect on the on-shell signal strength would increase for smaller mass values of the additional scalar (for example, for $m_S = 100\;\text{GeV}$ it reaches about
$50\%$).
As before, a weaker
upper-bound on $\kappa^2$ 
than for the case where the additional scalar in the loop is absent can be interpreted as a weaker
bound on the Higgs width ratio.
However, in view of the quite
small deviation of $\kappa^2$ from the off-shell bound that we find in  Fig.~\ref{fig:portallims}
no significant effect on the determination of the total width of the Higgs boson is expected in this case.

Instead of an additional scalar in the Higgs-boson propagator one could also consider the contribution of a spin-$1/2$ fermion or a spin-$1$ boson in the Higgs-boson propagator.
However, in simple cases of 
a single
state entering the Higgs-boson propagator one would not expect more sizeable contributions than for the singlet scalar case. An additional fermion that is coupled to the Higgs boson similarly to SM fermions would be excluded by fourth-generation fermion constraints. Instead, a vector-like fermion coupled to the Higgs boson and to a third-generation SM fermion would 
be less constrained. For example, a singlet vector-like lepton with 
a relatively
low mass is allowed by the present bounds from non-resonant Drell-Yan 
production~\cite{CMS:2022nty}. Assuming such a vector-like lepton with hypercharge $-1$ would couple to the Higgs doublet and the leptons via $ y L_L \Phi E_R + \text{ h.c.}$, electroweak precision observables would place strict limits on the mixing between the leptons~\cite{deBlas:2013gla}. For simplicity we consider only couplings of the vector-like lepton to $\tau$-leptons and calculate the renormalised Higgs self-energy. We find that for relatively low masses (below $1$~TeV) which would modify the off-shell $gg \rightarrow ZZ$ region, the renormalised self energy is much smaller than for the singlet scalar case~\cite{deBlas:2013gla}. Thus, such a scenario is not expected to have a larger impact on the Higgs width 
determination
compared to the Higgs portal scenario. The addition of a BSM vector boson $V_\mu$ coupled to the Higgs similarly to the SM vector bosons $\sim (H + v)^2 V_\mu V^\mu$ would lead to similar expressions as for the singlet scalar case; it is thus not expected to have a significant impact on the Higgs width determination if it is weakly coupled to the Higgs boson. 
Strongly coupled theories or the presence of multiple additional states (e.g.\ two vector-like fermions in the Higgs-boson self-energy) could however lead to different effects. We leave a more detailed discussion of the allowed parameter space and the possible size of the effects in scenarios of this kind for future work.

\section{Discussion and conclusions}
\label{sec:conc}

The total width of the detected Higgs boson at 125~GeV is a  crucial quantity for the physics of electroweak symmetry breaking
which
enters phenomenological studies of BSM physics in many ways. In particular, its value affects measurements of Higgs coupling modifiers which are commonly used as indirect constraints on new-physics scenarios.
The extraction of the Higgs coupling modifiers from the measured Higgs-boson signal strengths is in fact limited by the uncertainty on the total Higgs-boson width.

The by far most precise experimental determination of the total Higgs-boson width at the LHC relies on the theoretical assumption of equal on-shell and off-shell coupling modifiers in the $gg \rightarrow ZZ$ channel. In this paper we have relaxed this assumption and investigated how much the upper bound on the total Higgs-boson width can be weakened in realistic BSM scenarios that are in accordance with all existing experimental and theoretical constraints.

As a first step of our analysis
we have investigated the impact of 
an additional 
singlet scalar 
contributing to the $gg \to ZZ$ process via its lowest-order propagator. Such a contribution can
induce a destructive interference effect 
in the off-shell process which would be needed to compensate the enhancement of the coupling modifiers entering the on-shell Higgs process that is required to make a value of the total width which is significantly higher than the SM prediction compatible with the measurements of the on-shell signal strengths.
We have demonstrated that the case where the actual value of $\GamRat$ significantly exceeds the value that is obtained 
under the assumption of equal on-shell and off-shell coupling modifiers can only be realised if the additional scalar is very light, with a mass of about 200--300~GeV.
While the production of this additional scalar would yield a resonant contribution, this could be located below the signal region of existing searches in the $ZZ$ final state. Taking into account the existing limits, our analysis has shown that an enhancement of $\GamRat$ of up to 40\% is possible compared to the determination relying on the theoretical assumption about the on-shell and off-shell coupling modifiers for the smallest values of the mass of the additional scalar. Clearly, this scenario is severely constrained by the existing Higgs search limits, and future searches 
for light additional Higgs bosons in this channel can further tighten the obtained upper bound on the total width of the detected Higgs boson.

The next type of scenario that we have investigated are
modifications to the gluon-fusion production process from loop contributions of additional particles. Contributions of this kind could also yield destructive effects in the off-shell $ZZ$ channel. 
In this scenario important constraints arise from the fact that the gluon-fusion and weak boson fusion production processes for the on-shell results are modified in different ways which leads to important restrictions from the requirement of the simultaneous compatibility of the on-shell and off-shell predictions with the experimental results. Also in this case an upper limit on the Higgs-boson width going significantly beyond the one obtained under the assumption of equal on-shell and off-shell coupling modifiers can only be realised in the region where the additional particle is very light, for this scenario close to 100~GeV, and couples in a sizeable way to the known particles. Such a light coloured particle is obviously tightly constrained by existing searches.

A further scenario that we have investigated are loop contributions of additional particles to the Higgs-boson propagator, where we have explicitly spelled out the case of an additional scalar particle in the loop. The effects only become sizeable if the additional scalar has
a relatively large coupling to the 
Higgs boson. We find the largest enhancement for a mass of the scalar of about 200~GeV, but even in this case the possible enhancement of the total width remains relatively small.

While it is obviously not possible to cover all possible aspects of BSM scenarios, the results of our investigations suggest that in realistic BSM scenarios that are compatible with all existing experimental and theoretical constraints there is only limited room for realising a value for the total width of the detected Higgs boson that is higher than the upper bound that is obtained in the experimental analyses by ATLAS and CMS utilising \refeqs{eq:muon} and (\ref{eq:muoff}) and assuming equal on-shell and off-shell coupling modifiers. 
We have demonstrated in this context which conditions need to be fulfilled such that a larger value of the total width of the Higgs boson can be compatible with the existing results for on-shell and off-shell Higgs boson production. We have shown in the investigated scenarios that this implies that the presence of rather  
light additional particles with relatively large couplings to the known particles is required. Scenarios of this kind, however, are tightly restricted by the existing limits from searches for additional particles as well as by the results for electroweak 
precision observables and by theoretical constraints. 
While for strongly interacting scenarios and for scenarios with multiple additional states it is difficult to make quantitative statements about how much larger values for the total Higgs-boson width can be compatible with the existing constraints, the conclusion that relatively light new physics with sizeable couplings to the known particles is required for a significant modification of the upper limit should be maintained also for those wider classes of scenarios. Thus, they will be affected by the existing search limits and other constraints in a similar way as the scenarios that we have discussed in detail.  
Based on 
these considerations
we regard it as not very plausible that in a realistic BSM scenario a value of the total Higgs-boson width could be realised that exceeds the 95\% C.L.\ upper limit obtained from the indirect determination relying on the theoretical assumption of equal on-shell and of-shell coupling modifiers by a factor of more than about 2. We therefore expect that an upper limit on the total width of the detected Higgs boson that is twice as large as the one obtained based on the theoretical assumption of equal on-shell and off-shell coupling modifiers can be applied quite safely for scenarios where this theoretical assumption does not hold but which are compatible with the existing experimental and theoretical constraints.

While we have focused our investigation of
off-shell Higgs measurements on the $gg \rightarrow H \rightarrow ZZ$ 
channel, which was the first process studied in this context and up to now has by far the largest impact on the determination of the Higgs-boson width,
interesting information may also be obtainable during the HL-LHC phase from processes with
different final states.
For example, the contribution to top-pair production 
that is mediated by the detected
Higgs boson depends predominantly on the off-shell top-quark coupling modifier  $\kappa_{t,\text{off}}$ avoiding any dependence on vector boson couplings.
While experimental sensitivity to the contribution of the detected Higgs boson in this channel has been demonstrated~\cite{CMS:2019art,CMS:2020djy,ATLAS:2025ciy},
this channel is known to 
be affected by large signal--background 
interference effects which complicate its interpretation~\cite{Bahl:2025you,Bahl:2026ama}.
The production of four top quarks also depends solely on the top-quark coupling, but yields low cross-section rates at current luminosities.
Also the process of di-Higgs production is potentially of interest in this context as it involves both an off-shell Higgs exchange contribution and the on-shell production of the two Higgs bosons with their subsequent decays. Because of the different structure of this process and the simultaneous appearance 
of off-shell and on-shell coupling modifiers one would not necessarily expect that a similar cancellation between coupling modifiers and an enhanced total Higgs-boson width could occur as in \refeq{eq:muon}.
However, the achievable precision in the di-Higgs channel is unlikely to be high enough 
to have a significant impact on the determination
of the Higgs-boson width. It should also be stressed that caution is required in such 
analyses because of
the high sensitivity of such channels to 
possible effects of
BSM physics (similarly to the $ZZ$ process). 
Nevertheless, it will be interesting to exploit these additional processes 
or gaining complementary information.

It should be noted that the results of our analysis do by no means invalidate the physics potential of future $e^+e^-$ Higgs factories for measuring the total Higgs width and the Higgs branching ratios with minimal model dependence and high precision.
While the precision reachable on the total Higgs-boson width via the indirect determination at the HL-LHC can be a limiting factor in constraining BSM scenarios (see e.g.\ \citere{Forslund:2023reu}),
future Higgs factories will be able to improve on the HL-LHC precision very substantially~\cite{LinearColliderVision:2025hlt,deBlas:2025gyz}.

\subsection*{Acknowledgements}
We 
thank Christoph Englert, Guilherme Guedes and Kateryna Radchenko for useful discussions. 
This work was supported by the Deutsche
Forschungsgemeinschaft under Germany’s Excellence Strategy EXC2121
“Quantum Universe” - 390833306 and 
has been partially funded by the Deutsche Forschungsgemeinschaft 
(DFG, German Research Foundation) - 491245950.
P.S.\ acknowledges support by the European Research Council (ERC)
under the European Union’s Horizon 2020 research and innovation programme (Grant agreement No.~949451). 

\bibliographystyle{apsrev4-1_custom}
\bibliography{main.bbl}
\end{document}